\documentclass{IEEEcsmag}
\pdfoutput = 1
\usepackage[colorlinks,urlcolor=blue,linkcolor=blue,citecolor=blue]{hyperref}
\usepackage{lipsum}
\usepackage{upmath}
\usepackage{xifthen}
\usepackage{xspace}

\newcommand{\EMAIL}[1]{\href{mailto:#1}{#1}}
\newcommand{\TODO}[1]{{%
    \color{magenta} %
    \textbf{TODO}\ifthenelse{\equal{#1}{}}{\xspace}{:~}#1 %
  }}

\jvol{XX}
\jnum{YY}
\paper{Z}
\jmonth{December}
\jname{IEEE Computer}
\pubyear{2020}

\begin{document}

\title{Continuous Open Source License Compliance}
\author{Simon Phipps}
\affil{Meshed Insights Ltd, United Kingdom}

\author{Stefano Zacchiroli}
\affil{Université de Paris and Inria, France}


\begin{abstract}
  In this article we consider the role of policy and process in open source
  usage and propose in-workflow automation as the best path to promoting
  compliance.
\end{abstract}


\maketitle

Freely-licensed open source software (FOSS) has become a key ingredient
in software implementation across all sectors and at all scales of
enterprise. Once from the realm of informal hacker collaboration enabled
by the Internet, it has evolved into a discipline in its own right.
Whether approached from the pragmatic angle of OSI-approved licensing or
from the ethical angle of user rights, the fundamental enabler of FOSS
is the ability of developers to reuse and adapt pre-existing code
without the need to seek specific rights-holder approval or negotiate
terms. In practice that means it's easy to start at the point of
innovation rather than needing to first build (or buy) the
well-understood preliminaries and platforms others have already
pioneered.

The fact the developer can jump straight to innovation without asking
for basic permissions can conceal the fact that FOSS needs managing just
like any other software components. It has dependencies that will need
to be monitored and updated when there are security exposures; it has a
license that grants rights in return for satisfying responsibilities; it
comes from a source that needs at least a little ongoing engagement. In
this regard, it is just like proprietary software, just with more
freedoms.

Of these three management needs, satisfying the responsibilities
associated with the (otherwise freely permissive) license is the most
pressing. FOSS can be used only as a result of the grant of licenses to
the copyrights and patents its creators have generously offered. Their
corresponding requirements thus need respecting if the license is to
remain current. The actions associated with this compliance also bring
other benefits, such as the creation of a comprehensive manifest for
each subsystem (allowing dependency tracking to be performed) and the
acknowledgement of the communities and individuals involved. As such, it
is both an essential risk management activity and a best practice to
keep your code up-to-date and your company in good standing with the
communities upon which you depend.

As McAffer~\cite{DBLP:journals/computer/McAffer19} explained in this column,
these tasks are best performed by a dedicated \textbf{Open Source Programme
  Office} (OSPO) in larger organisations. But that doesn't mean it all has to
be a matter of manual effort! Increasingly, organisations are weaving license
compliance into the systems that also continuously build, test and deploy their
software. This serves multiple needs:

\begin{itemize}
\item
  \begin{quote}
  The routine hygiene of software for internal use
  \end{quote}
\item
  \begin{quote}
  The more consequential management of the manifest for software shipped
  to third parties or embedded in systems
  \end{quote}
\item
  \begin{quote}
  The specific needs that arise when a company is being prepared to
  acquire or be acquired.
  \end{quote}
\end{itemize}

While some commercial suppliers tend to emphasise the supposed risks
associated with the GNU General Public License (GPL) family, compliance
is much broader and more positive than that lens implies. The GPL does
indeed require licensees to make the Complete and Corresponding Source
code (CCS) available to recipients of the executable code that arises
from the use of their licensed materials, but almost every FOSS license
has requirements that it's important to respect. Licenses like BSD, MIT
and Apache all require a record of the previous contributors to be
passed on to users; licenses like MPL and EPL require certain
deliverables to be accompanied by source code to varying extents; and so
on.

These requirements are all a known quantity with current compliance
strategies and most are highly amenable to automation. Doing so controls
risk, improves quality and cultivates community respect. In this
article, we will thus focus on the current tools and popular workflows
that allow FOSS license compliance to be satisfied invisibly most of the
time.

\section{THE OPEN SOURCE SUPPLY CHAIN}

Given almost every enterprise software system contains open source software,
where has it come from? How is it manipulated and assembled to produce internal
systems? Who receives the resulting constructions?  These questions define a
\textbf{software supply chain}, which may involve a surprisingly long and broad
sequence of entities on the inbound side and could include third parties on the
outbound side even if your business does not apparently trade in software. A
previous column by Harutyunyan~\cite{DBLP:journals/computer/Harutyunyan20}
covers this concept in more depth.

\begin{figure*}
  \centering
  \includegraphics[width=\textwidth]{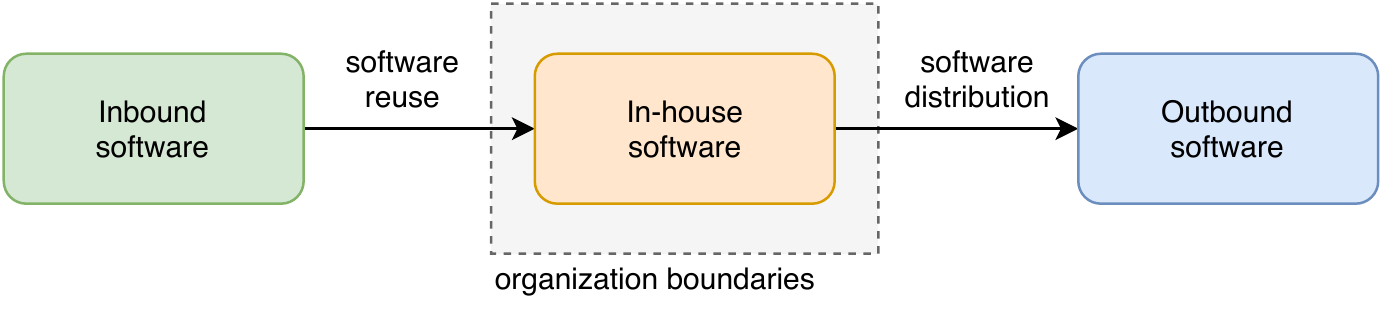}
\end{figure*}

In brief, your open source supply chain comprises \emph{inbound
software}---the open source \emph{and} proprietary software entering
your enterprise---together with its dependencies; \emph{in-house
development,} the adaptations you make to inbound software and the
software you develop yourself; and then \emph{outbound software,} the
software you pass to others either under open source licenses or under
proprietary terms, either as software, as software-implemented services
or embedded in hardware.

Managing your open source supply chain will require a comprehensive
\textbf{open source policy}. Fundamentally, your open source policy
encapsulates the risks that you consider justified within your business.
This will include a determination of which licenses you have analysed
and understood, which combinations of those licenses are acceptable
within both inbound and outbound software, what management steps are
required to ensure the risks are managed and how the determinations are
recorded and reviewed.

Your open source policy should go beyond licensing however. You will
need policies on how vulnerabilities discovered in inbound software will
be evaluated and what steps will be taken to ensure all uses of the
affected inbound software are addressed. It is important to have
policies on what skills are required in-house for managing critical
components and how staff absences/departures will be handled, for
example through third-party contractors or through supplier
subscriptions. You should also have a policy and associated budget for
memberships and sponsorships of community charities (OSI, FSF, Apache
etc.) and software trade associations (Eclipse Foundation, Linux
Foundation and so on) so that you are able to properly exercise
influence as well as make upstream contributions. McAffer has covered
governance topics and open source policies in more depth in this series.

These policies will then drive your \textbf{open source review process}.  This
will be triggered when someone wishes to use inbound software that has not
previously been evaluated. Your review will address the needs detailed in your
open source policy---including licensing, staffing, maintenance, reputational
issues, community influence, component maturity and more. A previous article by
Spinellis~\cite{DBLP:journals/computer/Spinellis19} has covered all the factors
that you should take into account when choosing a FOSS component.

At the conclusion of the review, the inbound software will be given a
go/no-go determination, usually by your OSPO. The result of the
assessment will be recorded so that future use of cleared components
does not require any further permission to proceed. This is essential if
the core value of open source is to be leveraged.

One further step is the creation of your policy for upstream
contribution. Your developers should be able to freely contribute their
improvements to the upstream maintainers of your inbound software. Your
policy and associated process should ensure that all contributor
agreements are reviewed and approved in advance, and that your patent
portfolio is not inadvertently used against upstream communities or
inbound software.

\section{COMPONENT INVENTORY MANAGEMENT}

Specific tools exist to support the business processes related to open
source reviews, in the form of \textbf{component inventory managers}.
Eclipse SW360
(\href{https://www.eclipse.org/sw360/}{{https://www.eclipse.org/sw360/}})
is a popular open source solution in that space and a cornerstone of
most enterprise FOSS governance workflows. Let's see how it works, as a
classic example of a tool supporting the inbound part of software supply
chains.

All software components, open source or otherwise, known to your
organization will be added to the organization-specific SW360 instance
and assigned a \emph{canonical} name, allowing to recognize components
carrying different names in different contexts (upstream repository,
distribution, package manager, etc.) as the same. The source code of
component \emph{releases} will also be uploaded to SW360 and from there
automatically analyzed with license scanners, such as FOSSology or
ScanCode. Crucially, the addition of novel FOSS components to SW360 can
trigger \emph{clearance requests}, that only specific users (e.g., OSPO
members) can perform before the component is deemed fit for use. Once a
component is cleared it will remain so for all future uses. Component
reviews and clearance decisions are typically based on license scanning
results, but can also take into account the other factors we have
discussed, such as, known vulnerabilities, development activity, bus
factor, etc.

SW360 can also maintain a mapping between the IT products distributed by
your organization and the FOSS components they contain, e.g., as
dependencies or reused code. This mapping enables SW360 to automatically
produce license compliance documents such as \textbf{Software Bills of
Material} (SBOM, cf. \emph{Managing Your Open Source Supply Chain},
previously on this column) in machine-readable standard formats such as
SPDX, as well as human-readable documents like the list of all FOSS
licenses used in the product, copyright notices for attribution
purposes, and offers for source relevant to a given product.

All workflows in SW360 can be performed manually via a web portal or
automated via command-line tools and RESTful APIs. You can for instance
further support the clearance process by plugging other scanners into
the scanning subsystem as well as integrate information retrieved from
external knowledge bases, e.g. ClearlyDefined or Software Heritage. The
reliance on SPDX as an exchange format allows SW360 to automate the
import of entire SBOMs for inbound software, as well as the production
of complete and corresponding source (CCS) tarballs for outbound
software, when required by the
license.

\section{CONTINUOUS LICENSE COMPLIANCE}

\begin{figure*}
  \centering
  \includegraphics[width=\textwidth]{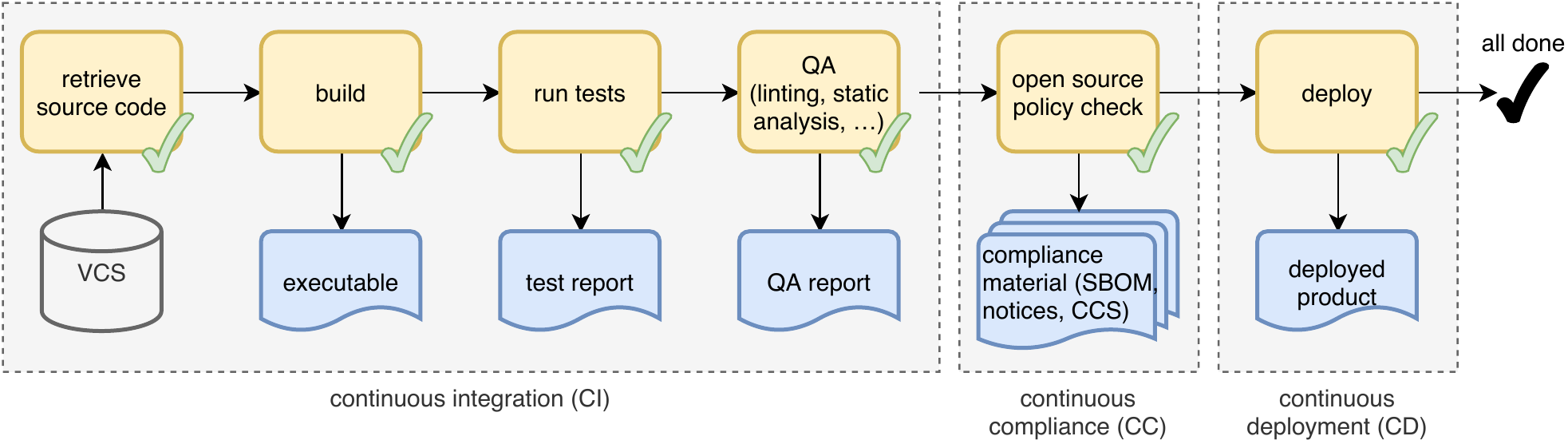}
\end{figure*}

Once you have a clear grasp of cleared components in your software
supply chain and can fulfill license obligations \emph{at a specific
point in time}, the question that naturally arises is how to maintain
that status quo \emph{in the long run}, as development continues. This
boils down to deciding \emph{when} you review your products for
adherence to your internal open source policy.

Various approaches are possible, including sporadic ``fire drills'',
periodic checks, and release-time checks. The best approach is to
\emph{continuously and automatically} check that all your products
adhere to your open source policy---this is what we refer to as
\textbf{continuous open source license compliance} in this article, or
\emph{continuous compliance} for short.

In practice this means integrating license compliance checks into
existing continuous integration/continuous deployment (CI/CD)
toolchains, \emph{making the build fail} in case a divergence between
the actual build artifacts and your open source policy is detected.
Doing so immediately notifies developers of policy issues that should be
dealt with, resulting in shorter feedback loops (a core tenet of agile
development practices) and reduced risks of wasted development efforts
(e.g., integration of a FOSS component that will need to be replaced
before release).

For continuous compliance to work, the verification of adherence to your
open source policy should be as automated as possible, so that it stays
out of the way of engineers---until things go wrong.

\section{TOOLCHAIN INTEGRATION}

Implementing in practice the general idea of continuous compliance as a
fully-automated verification of adherence to your open source policy can
be tricky. No one-size-fits-all solution is well-established in the
industry yet, nor will probably ever be, due to how each build toolchain
is a unique snowflake. Your own implementation of continuous compliance
will likely combine several tools and integrate them with custom
adapters.

This need is behind the current industry push toward \textbf{open
compliance}, i.e., the idea that one should preferably rely on FOSS
tools to implement continuous compliance. With FOSS compliance tools it
is easier to tailor tools to your specific needs, lock-in risks get
mitigated, and you will have opportunities to contribute back to the
ecosystem, remaining on top of the technology you use and maintaining
influence on software evolution. Proprietary compliance tools can still
be used, but are usually integrated as black boxes into toolchains that
contain increasingly large majorities of open tools.

The tooling landscape
(\href{https://github.com/Open-Source-Compliance/Sharing-creates-value/tree/master/Tooling-Landscape}{{https://github.com/Open-Source-Compliance/Sharing-creates-value/}})
conducted by the Open Source Tooling Group and the OpenChain curriculum
(\href{https://github.com/OpenChain-Project/curriculum/raw/master/slides/openchain-curriculum-for-2-0.pdf}{{https://github.com/OpenChain-Project/curriculum/}})
provide a good overview of existing tools to support automated
governance of FOSS supply chains. As there are too many to be explained
here in full, we conclude with a few examples and discuss how they can
be used to implement continuous compliance.

Several high-quality \textbf{license and source code scanners} exist. As we
have seen, their integration into continuous compliance happens at the
component inventory level, during clearance of inbound components.  We refer to
the comprehensive overview of such tools given by
Ombredanne~\cite{DBLP:journals/computer/Ombredanne20} last month on this column
for more details.

\textbf{Dependency trackers} ensure that all the dependencies pulled in
at build time are known and cleared in your component inventory. This is
of paramount importance because, while you can easily audit the
\emph{direct} dependencies declared in your software products,
\emph{transitive dependencies} can change in the ecosystem unexpectedly
and might pull in unknown components or newer versions of known
components yet to be cleared.

\emph{Eclipse SW360 Antenna}
(\href{https://www.eclipse.org/antenna/}{{https://www.eclipse.org/antenna/}})
is a popular tool used to automate dependency tracking that, in
conjunction with SW360, constitutes a fairly comprehensive continuous
compliance implementation. Antenna integrates with build automation
tools, like Maven or Gradle, scans source code artifacts of all
dependencies during build, verifies that they are cleared in SW360---or,
alternatively, populates it with them so that they can be cleared later
on---and automatically produce compliance material such as the list of
all dependencies with their licenses in a spreadsheet-friendly format
and a CCS bundle containing all their source code. Integration with
other build tools is possible with custom code and the build can be made
to fail in case non-cleared components are encountered or if their
license does not adhere to your open source policy.

The overall continuous compliance ecosystem is moving at a fast pace,
with new tools and approaches being frequently released. Other
noteworthy players in the same feature space of what SW360 and Antenna
offer are Quartermaster
(\href{https://qmstr.org/}{{https://qmstr.org/}}), hosted by The Linux
Foundation as part of their Automated Compliance Tooling (ACT) umbrella
project, and the OSS Review Toolkit (ORT,
\href{https://github.com/oss-review-toolkit/ort}{{https://github.com/oss-review-toolkit/ort}}).
The latter is particularly interesting as it provides a highly
customizable pipeline for continuous compliance, composed of several
independent blocks: a dependency analyzer, a downloader for dependencies
source code, an abstraction over license scanners, a policy checker that
supports custom business rules, and a reporter to build SBOMs. ORT
components can, and often are, used independently from the other blocks
in custom continuous compliance toolchains.

\section{CONCLUSION}

Continuous open source license compliance is now a well-established
industry best practice in managing the life cycle of software products.
It consists of automating as much as possible the verification of the
adherence of your IT products, that almost invariably contain open
source components, to the open source policy of your organization.
Ideally, such an automated verification is then integrated into your
existing CI/CD toolchain, making software builds fail and notifying
developers early when issues are spotted.

No one-size-fits-all technology to implement continuous compliance has
emerged yet, due to the heterogeneity of build toolchains. On the other
hand there is consensus on the \emph{types} of tools you will need:
component inventories, scanners, dependency trackers, policy checkers,
and generators of compliance material (SBOMs, notices, and CCS bundles)
are all tools of the trade. High quality open source implementations of
all these tools exist and should be used as the basis for addressing
your specific continuous compliance needs.

\begin{IEEEbiography}{Simon Phipps}{\,}%
  leads the management consultancy he founded, Meshed Insights Ltd, and among
  other industry roles was head of one of the first staffed OSPOs (at Sun
  Microsystems). Contact him at: \EMAIL{simon@meshedinsights.com}.
\end{IEEEbiography}

\begin{IEEEbiography}{Stefano Zacchiroli}{\,}%
  is associate professor of Computer Science at Université de Paris, on leave
  at Inria. He is the co-founder and CTO of the Software Heritage project.
  Contact him at \EMAIL{zack@irif.fr}.
\end{IEEEbiography}



\begin{thebibliography}{1}

\bibitem{DBLP:journals/computer/Harutyunyan20}
Nikolay Harutyunyan.
\newblock Managing your open source supply chain-why and how?
\newblock {\em Computer}, 53(6):77--81, 2020.

\bibitem{DBLP:journals/computer/McAffer19}
Jeff McAffer.
\newblock Getting started with open source governance.
\newblock {\em Computer}, 52(10):92--96, 2019.

\bibitem{DBLP:journals/computer/Ombredanne20}
Philippe Ombredanne.
\newblock Free and open source software license compliance: Tools for software
  composition analysis.
\newblock {\em Computer}, 53(10):105--109, 2020.

\bibitem{DBLP:journals/computer/Spinellis19}
Diomidis Spinellis.
\newblock How to select open source components.
\newblock {\em Computer}, 52(12):103--106, 2019.

\end{thebibliography}
\end{document}